\documentclass[twocolumn,showpacs,preprintnumbers,amsmath,amssymb,pre]{revtex4-1}

\usepackage{color}    
\usepackage{graphicx}
\usepackage{dcolumn}
\usepackage{bm}
\usepackage{subfigure}
\usepackage{amssymb}
\usepackage{multirow}
\usepackage{amsmath}
\usepackage{array}
\graphicspath{{plots/}}

\renewcommand{\vec}[1]{\mathbf{#1}}

\begin{document}

\title{Dynamical structure factor of strongly coupled ions in a dense quantum plasma}

\author{ Zh.~A.  Moldabekov$^{1, 2}$, H.~K\"ahlert$^{1}$, T.~Dornheim$^{1}$, S.~ Groth$^{1}$,   M. Bonitz$^{1}$, and T. S.  Ramazanov$^{2}$}

\affiliation{
 $^1$Institut f\"ur Theoretische Physik und Astrophysik, Christian-Albrechts-Universit\"at zu Kiel,
 Leibnizstra{\ss}e 15, 24098 Kiel, Germany}
 \affiliation{
 $^2$Institute for Experimental and Theoretical Physics, Al-Farabi Kazakh National University, 71 Al-Farabi str.,  
  050040 Almaty, Kazakhstan
  }
\begin{abstract}
The dynamical structure factor (DSF) of strongly coupled ions in dense plasmas with partially and strongly degenerate  electrons is investigated. The main focus is on the impact of electronic correlations (non-ideality) on the ionic DSF.  The latter is computed by carrying out molecular dynamics (MD) simulations with a screened ion-ion interaction potential. The electronic screening is taken into account by invoking the Singwi-Tosi-Land-Sj\"olander approximation, and compared to the MD simulation data obtained considering the electronic screening in the random phase approximation and using the Yukawa potential. 
We find that electronic correlations lead to lower values of the ion-acoustic mode frequencies and to an extension of the applicability limit with respect to the wave-number of a hydrodynamic description. Moreover, we show that even in the limit of weak electronic coupling, electronic correlations have a non-negligible impact on the ionic longitudinal sound speed. Additionally, the applicability of the Yukawa potential with an adjustable screening parameter is discussed, which will be of interest, e.g., for the interpretation of experimental results for the ionic DSF of dense plasmas.

\end{abstract}

\pacs{xxx}

\maketitle
\section{Introduction}

Dense plasmas with partially or strongly degenerate electrons and non-ideal ions are nowadays routinely realized in direct-drive ignition at
the NIF \cite{Hu, Paisner}, cryogenic DT implosion on OMEGA \cite{Hu, Boehly}, and other intense heavy ion and laser beams experiments \cite{ Hoffmann1, Sharkov, Glenzer, Ma, Lee}. Additionally to inertial confinement fusion, these experiments with matter under extreme conditions are motivated  by the need to understand astrophysical objects like giant planets and white dwarfs.
Of particular interest in this context is the dynamic structure factor (DSF) $S(k,\omega)$, which is a direct measure for the response of a system to an external perturbation and is an important tool for diagnostics \cite{Saunders}.
Previously, the DSF of the ions was extensively investigated using Coulomb and Yukawa potentials as pair interaction \cite{Hansen, Donko, Mithen, Mithen2}, which are referred to as the one-component Coulomb plasma model (OCP) and the Yukawa one-component plasma model (YOCP), respectively. 

Recent advances in the field of inelastic X-ray scattering \cite{McBride} provide the means to measure the dynamical structure factor of ions, which allows to experimentally probe the ion-acoustic mode, sound velocity and transport coefficients. Consequently, this improvement in diagnostic capabilities has triggered a spark of theoretical investigation of the ionic DSF over the last few years~\cite{White, Rueter, Vorberger, Mabey}.

At warm dense matter (WDM) conditions (see, e.g., Ref.~\cite{dornheim_pr_2018}) for a topical overview and concise definition), the importance of the short-range repulsion between ions was indicated by White et al. \cite{White} by analyzing the ionic structural features [performing orbital-free density functional molecular dynamics simulation (OFMD)] and by Vorberger et al. \cite{Vorberger} (by analyzing Kohn-Sham density functional theory (DFT) simulation results).  Later, this aspect was carefully revised by Cl\'erouin et al. \cite{Clerouin} and Harbour et al. \cite{Harbour} by investigating a two-temperature WDM system on the basis of OFDFT and an improved neutral-pseudoatom model, respectively. They found that the ionic structural features, which were interpreted to be the result of the short-range ion-ion repulsion, can be the result of a nonisothermal (out-of-equilibrium) state of WDM, which was not accounted for in the previous studies.  Additionally, it was shown that the strong ionic coupling leads to the requirement of a more accurate treatment of the electronic quantum and correlation effects~\cite{PRE18}.   Moreover, recently Mabey et al. \cite{Mabey} showed that dissipative processes (which were included by a friction term in the utilized Langevin dynamics) can have a strong impact on the ionic DSF. Therefore, 
the understanding of the ionic DSF in dense quantum plasmas and in WDM depends upon an accurate analysis of the impact of various factors (quantum degeneracy, electronic correlations, the effects beyond the Born-Oppenheimer approximation etc.). Without  doubt, the availability of well-defined models with clear approximations are invaluable for such an analysis. Indeed, the comparison of density functional theory simulation results to the OCP and YOCP served as a valuable tool for the understanding of the static and dynamic properties of WDM \cite{Clerouin, Clerouin2, Wunsch}. 

In this work we focus on the role (impact) of electronic correlations (non-ideality) on the ionic DSF in dense plasmas with partially or strongly degenerate electrons and strongly coupled ions. The investigation is carried out by analyzing the data from molecular dynamics (MD) simulations of ions \cite{ludwig_jpcs_10} where the electronic screening is taken into account by invoking the well-known Singwi-Tosi-Land-Sj\"olander approximation (STLS) \cite{stlsT0, stls}. In particular, the STLS approximation allows us to take into account electronic exchange-correlation effects \cite{dornheim_pr_2018}. The information about the impact of electronic correlations on the ionic DSF is subsequently extracted by comparing the STLS-based results to the data from MD simulations with the screening being described on the level of the random phase approximation (RPA, i.e., by neglecting the electronic exchange-correlation effects). Additionally, we discuss the applicability of a Yukawa model for the description  of the ionic DSF, which will be of interest, e.g., for the interpretation of experimental data or the results from ab-initio simulations such as DFT.

The considered plasma parameters and a description of the used screened ion potentials are given in Sec.~\ref{s:2} and Sec.~\ref{s:3}, respectively. 
The simulation results are presented in Sec.~\ref{s:4}.

\section{Plasma parameters}\label{s:2} 

The electronic component of dense plasmas is conveniently characterized by the density $r_s=a_e/a_B$ and degeneracy parameters $\theta=k_B T_e/E_F$, where $a_e=(3/4 \pi n_e)^{-1/3}$ , $E_F$ is the Fermi energy of electrons, $a_B$ denotes the first Bohr radius, and $n_e$ ($T_e$) is the electronic number density (temperature). In terms of $r_s$ and $\theta$, the temperature and density of electrons is expressed as $T\simeq (\theta/r_s^2)\times 0.58\times 10^{6}~{\rm K}$ and $n_e\simeq 1.6\times 10^{24}~r_s^{-3}$, respectively.
The classical ionic component is described by the coupling parameter $\Gamma=Z_i^2e^2/(ak_BT_i)$, where $a=(3/4 \pi n_i)^{-1/3}$ is the mean inter-ionic distance, $T_i$ is the temperature of ions, and $n_i=n_e/Z_i$.  The temperature of ions and electrons can be different.  A transient non-isothermal state (with $T_i\neq T_e$) appears due to slow electron-ion temperature relaxation rate resulting from the large ion-to-electron mass ratio \cite{Hartley, White2014, MRE2017, Gericke}. 

We consider dense plasmas with $r_s\leq 1.5$ and $\theta\lesssim 1$. 
To be more precise, the simulations have been carried out for $r_s=1.5$, $r_s=1.0$, and $r_s=0.5$ in the range of degeneracy parameters from $\theta=0.1$ to $\theta=2.0$. These values of the parameters are chosen to track the impact of electronic correlations starting from the case of non-ideal electrons at $r_s=1.5$ to the case of weakly coupled electrons at $r_s=0.5$. Additionally, electronic thermal excitation effects can be measured by comparing the strongly degenerate case at $\theta=0.1$ to the semi-classical regime, $\theta=2.0$.    

The corresponding range of the temperature and density of the electrons is $2.6\times 10^{4}~{\rm K}\leq T_e\leq 4.6\times 10^{6}~{\rm K}$ and $5\times 10^{23}~{\rm cm^{-3}}\leq n_e \leq 1.3\times 10^{25}~{\rm cm^{-3}}$, respectively.
At these parameters, the plasma is fully or highly ionized due to thermal and pressure ionization \cite{Chabrier, Bonitz2005}.

The plasma parameters that are considered in this work have already been realized experimentally
during cryogenic DT implosion on OMEGA, direct-drive ignition
at the NIF \cite{Hu, Boehly} as well as in other laser-driven shock-compression experiments \cite{Ma, Glenzer, Lee}.  

Further, for simplicity and without loss of generality, we take $Z_i=1$ and $\Gamma=15$. The corresponding temperature ratio is given by $T_e/T_i\simeq 1.84 \times (\theta/r_s)\Gamma$ and considered to be $T_e/T_i\lesssim 20$ in accordance with experiments (see discussions in Refs.~\cite{PRE18, Clerouin, Harbour, Plagemann}). For example, at $r_s=0.5$ and $\theta=0.1$ we have $T_e/T_i\simeq 5.5$ and at $r_s=1.5$ and $\theta=1$ we find $T_e/T_i\simeq 18$.

\section{Screened potentials and simulation details} \label{s:3} 

At the plasma parameters that are of interest in the context of the present work, the weak electron-ion
coupling allows to decouple the electronic and ionic dynamics by introducing  the
 screened ion potential \cite{PRE18, Hansen}:
\begin{multline}\label{eff_pot_1}
 \Phi(\vec r_{j^{\prime}},\vec r_j)= \frac{Z_i^2e^2}{|\vec r_{j^{\prime}}-\vec r_j|}+\\ \int\! \frac{\mathrm{d}^3k}{(2 \pi)^3 } \left| \tilde \varphi_{\rm ei}(\vec k)\right|^2  \chi_e(\vec k)  \,\, e^{i \vec k \cdot (\vec r_{j^{\prime}}-\vec r_j)} \quad,
\end{multline}
where $\tilde \varphi_{\rm ei}(\vec k)=4\pi Z_ie^2/k^2$ is the bare electron-ion Coulomb interaction potential, and the screening by the electrons is described by the density response function from linear response theory,
\begin{equation}\label{chi_G}
\chi_e^{-1}(\vec k, \omega)=\chi_0^{-1}(\vec k, \omega)+\frac{4\pi e^2}{k^2}\left[G(\vec k, \omega)-1\right].
\end{equation}
In Eq.~(\ref{chi_G}), the electronic exchange-correlation effects are taken into account in the local field correction $G(\vec k, \omega)$. If the electronic non-ideality is neglected ($G=0$), Eq.~(\ref{chi_G}) simplifies to the  density response function of electrons  in RPA,  with  $\chi_0$ being the ideal density response function  at a finite temperature \cite{quantum_theory}.

A widely used and successful approximation for the static local field correction is the self-consistent static STLS ansatz ~\cite{stlsT0,stls}.
Within the STLS scheme, the static local field correction is computed by solving the following system of non-linear equations:
\begin{equation}
G^\textnormal{STLS}(\mathbf{k},0) = -\frac{1}{n} \int\frac{\textnormal{d}\mathbf{k}^\prime}{(2\pi)^3}
\frac{\mathbf{k}\cdot\mathbf{k}^\prime}{k^{\prime 2}} [S_e^{\text{STLS}}(\mathbf{k}-\mathbf{k}^\prime)-1]\;,
\label{G_stls}
\end{equation}
and 
\begin{equation}\label{flucDis}
 S_e^\text{STLS}(\mathbf{k}) = -\frac{1}{\beta n}\sum_{l=-\infty}^{\infty} \frac{k^2}{4\pi e^2}\left(\frac{1}{\epsilon(\mathbf{k},z_l)}-1\right),
\end{equation}
where $S_e^{\textnormal{STLS}}$  is the static structure factor of the electrons. Note that the summation in Eq.~(\ref{G_stls}) is taken over the so-called Matsubara frequencies, $z_l=2\pi il/\beta\hbar$. 

In the following, the potential computed using Eqs. (\ref{eff_pot_1})--(\ref{flucDis}) is referred to as the \textit{STLS potential}.
For completeness, we mention that Dornheim, Groth, and co-workers ~\cite{dornheim_pr_2018, Dornheim_PRL18, Dornheim_PRE17, Groth17} have recently obtained the first exact data for $G(q,\omega=0)$ at finite temperature on the basis of ab-initio quantum Monte Carlo (QMC) simulations. However, at the time of writing, these data are available only for a few parameters, and a detailed comparison between the STLS and QMC results, and their respective impact on the ionic DSF, remains an interesting topic for future research.

To reveal the importance of the electronic non-ideality for the dynamics of strongly coupled ions, we also use the \textit{RPA potential} calculated by neglecting electronic correlations in the density response, i.e., taking $G=0$. The STLS and RPA potentials are then used for the computation of the dynamic structure factor via MD simulations. 

In our previous works \cite{CPP17, PRE18}, we have already provided a detailed discussion of the STLS potential and compared it to potentials that were obtained using RPA. The main feature of the STLS potential is that electronic correlations lead to a stronger screening of the ion charge in comparison with the RPA potential. Additionally, in Ref.~\cite{PRE18} we have shown the applicability range of the STLS approximation for the computation of screening with respect to $r_s$ and $\theta$ by comparing it to ground state QMC data in the low temperature limit.

Here, we perform MD simulations for $N=6400$ ions within periodic boundary conditions. The equilibration time following random initialization of coordinates and velocities was $T_{eq}=1000~\omega_{pi}^{-1}$, where $\omega_{pi}$ is the ionic plasma frequency. After the equilibration time, the computation of the dynamical structure factor was performed during a  $7600~\omega_{pi}^{-1}$  time period with the time step $dt=0.02~\omega_{pi}^{-1}$.  The final result for DSF was obtained after averaging over 40 independent simulations for each set of plasma parameters.    

\begin{figure}[h!]
  \vspace{0.5cm}
\includegraphics[width=0.49\textwidth]{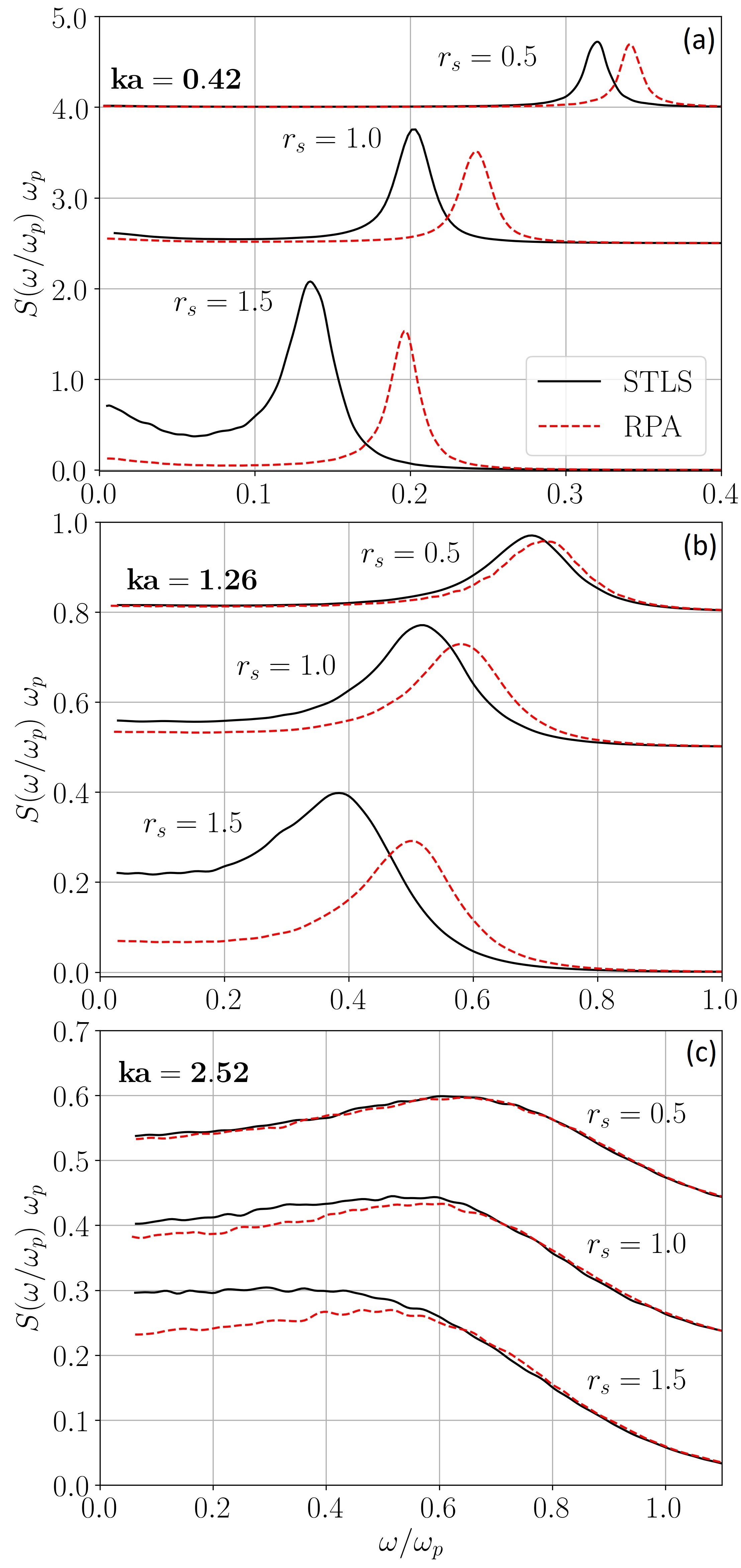}
\caption{Dynamical structure factor of ions at $\theta=0.1$ and at three values of the density parameter. The data obtained using STLS and RPA potentials is denoted as STLS and RPA, respectively. The DSF curves for $r_s=1.5, \,1.0$ and $r_s=0.5$ are shifted vertically for clarity.}
\label{fig:1}
\end{figure}

\section{Simulation results}\label{s:4} 
\subsection{Results for STLS and RPA potentials}\label{s:4A}
First we consider the dependence of the ionic DSF on $r_s$ and $\theta$.
In Fig.~\ref{fig:1}, the DSF is shown at $\theta=0.1$ and densities $r_s=1.5$, $r_s=1$, and $r_s=0.5$. In this figure, the dependence of the DSF on $\omega$ is given for (a) $ka=0.42$, (b) $ka=1.26$, and (c) $ka=2.52$. From the comparison of the STLS potential based results and the RPA potential based results in Fig.~\ref{fig:1}, we see that the electronic correlations lead to a shift of the maximum to lower frequencies  (red-shift). Additionally, electronic correlations result in a higher value of both the maximum in the DSF and $S(k,0)$. While the former represents the ion-acoustic mode, the latter is due to diffusive thermal motion and referred to as the Rayleigh peak  \cite{Hansen_book} . 
To provide more accurate information about the mentioned differences, in Table~\ref{tab:1} we compare the maximum positions and heights for the STLS-potential- and RPA-potential-based results at $ka=0.42$. With increasing density, the difference in the position of the ion-acoustic peak decreases from $30~\%$ at $r_s=1.5$ to $17~\%$ at $r_s=1.0$ and attains a value of $6~\%$ at $r_s=0.5$. The difference in height of the ion-acoustic peak is $29.6~\%$ at $r_s=1.5$, $20.5~\%$ at $r_s=1.0$, and reduces to $3~\%$ with a further increase in density to $r_s=0.5$. With increasing wave-number, the differences between the STLS and RPA data become less pronounced [see Figs. ~\ref{fig:1} (b) and (c)]. Note that the increase in wave-number is accompanied by a decrease of the ion-acoustic peak, which shifts to higher frequencies and becomes significantly broader. At $ka=2.52$, the ion-acoustic peak almost disappears, see Fig.~\ref{fig:1}(c).
The Rayleigh peak also disappears for larger wave-numbers. There is no longer a prominent maximum at $\omega=0$, already at $ka=1.26$, but $S(k, 0)\neq 0$.

\begin{figure}[h!]
  \vspace{0.6cm}
\includegraphics[width=0.49\textwidth]{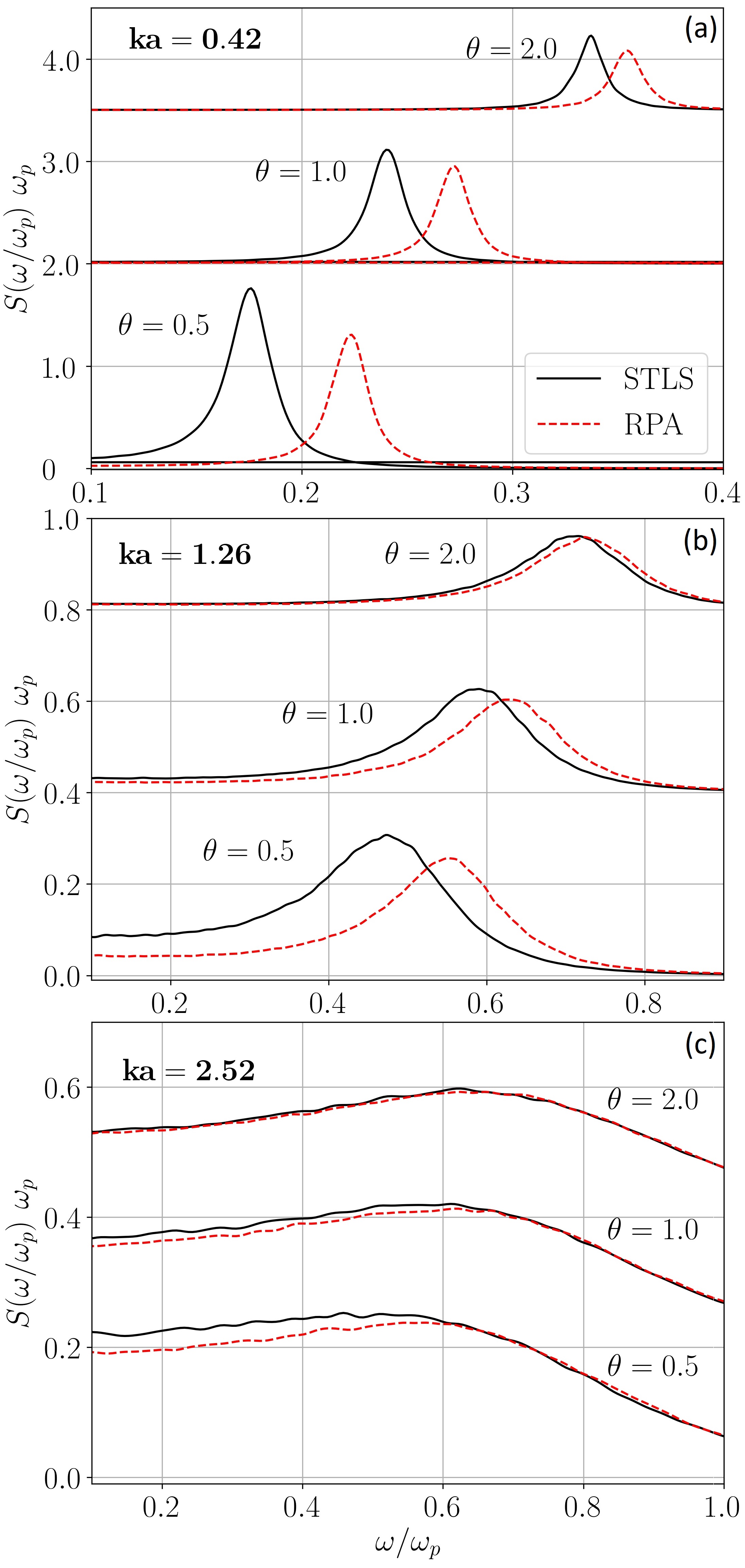}
\caption{Dynamical structure factor of ions at $r_s=1.5$ and at different values of the degeneracy parameter. The data obtained using the STLS and RPA potential are denoted as STLS and RPA, respectively. The DSF curves for  $\theta=1$ and  $\theta=2$ are shifted for clarity. }
\label{fig:2}
\end{figure}

Now we investigate the behavior of the DSF of the ions going from the strongly degenerate case, $\theta=0.5$, to the partially degenerate case with $\theta=1$ and $\theta=2$. The results for these values of $\theta$ at $r_s=1.5$ are shown in Fig.~\ref{fig:2}. Again, the DSF curves are given for (a) $ka=0.42$, (b) $ka=1.26$, and (c) $ka=2.52$. In Fig.~\ref{fig:2}, we clearly see that thermal excitations suppress the  influence of electronic correlations on the ionic DSF. In Table \ref{tab:2}, we provide numbers for differences in positions and heights of STLS potential and RPA potential based data at $ka=0.42$. In the case of $r_s=1.5$, the difference in the ion-acoustic peak position and height of the STLS potential based result and that of RPA potential based result is $18~\%$ and $35.7~\%$, respectively. The difference in the ion-acoustic peak position reduces to $11~\%$ at $\theta=1.0$ and further to $3~\%$ at $\theta=2.0$. Similarly, the difference in height decreases to $16~\%$ at $\theta=1.0$ and to $9~\%$ at $\theta=2.0$. At larger values of the wave-number, the differences between the STLS potential based results and RPA potential based results become smaller (see Figs.~\ref{fig:2} (b) and (c)). For example, at $ka=1.26$ and $\theta=2.0$, the electronic correlations can be safely neglected.

We note that, in the limit of large  wave-numbers, the DSF should eventually approach the free-particle limit, which has a Gaussian shape~\cite{Mithen}. In this limit, the deviations between different potentials are expected to vanish.

\begin{table}[h]
      \centering
          \caption{The acoustic mode frequency at $ka=0.42$ for $\theta=0.1$, where $\omega_{\rm max}$ is the position of the ion-acoustic peak and $S^*=S(\omega_{\rm max})\omega_p$,   $\Delta \omega=\omega_{\rm max}^{\rm RPA}-\omega_{\rm max}^{\rm STLS}$. In all cases the width of DSF at the half of its maximum height is $0.2~\omega_p$.}\label{tab:1}
      \vspace{0.5cm}
      \begin{tabular}{|c| c| c| c|c|}\hline
          $r_s$ &  \multicolumn{2}{|c|}{\text{$\omega_{\rm max}/\omega_p$}} &  \text{ $\frac{\Delta \omega}{\omega_{\rm RPA}}\times 100~\%$} & \text{$\frac{S_{\rm STLS}^*-S_{\rm RPA}^*}{S_{\rm RPA}^*}\times 100~\%$}   \\
           \cline{2-3}
           & \text{STLS}& \text{RPA}&&\\
           \hline\hline
        1.5& 0.14 & 0.2 & 30~\%& 29.6~\%\\\hline
         1.0& 0.2 & 0.24 & 17~\% & 20.5~\% \\\hline
            0.5& 0.32 & 0.34 & 6~\%& 3~\% \\\hline
      \end{tabular}
      \end{table}
      
      \begin{table}[h]
      \centering
          \caption{The acoustic mode frequency at $ka=0.42$ for $r_s=1.5$, where $\omega_{\rm max}$ is the position of ion-acoustic peak and $S^*=S(\omega_{\rm max})\omega_p$,    $\Delta \omega=\omega_{\rm max}^{\rm RPA}-\omega_{\rm max}^{\rm STLS}$. At $\theta=0.5$ and $\theta=1.0$, the width of DSF at the half of its maximum height is $0.2~\omega_p$. At $\theta=2.0$, the width of DSF is $0.14 ~\omega_p$ for STLS and $0.16 ~\omega_p$ for RPA. }\label{tab:2}
      \vspace{0.5cm}
      \begin{tabular}{|c| c| c| c|c|}\hline
          $\theta$ &  \multicolumn{2}{|c|}{\text{$\omega_{\rm max}/\omega_p$}} & \text{$\frac{\Delta \omega}{\omega_{\rm RPA}}\times 100~\%$} & \text{$\frac{S_{\rm STLS}^*-S_{\rm RPA}^*}{S_{\rm RPA}^*}\times 100~\%$}   \\
           \cline{2-3}
           & \text{STLS}& \text{RPA}&&\\
           \hline\hline
        0.5& 0.18 & 0.22 & 18~\%& 35.7~\%\\\hline
         1.0& 0.24 & 0.27 & 11~\% & 16~\% \\\hline
            2.0& 0.34 & 0.35 & 3~\%& 9~\% \\\hline
      \end{tabular}
      \end{table}

To achieve a more detailed understanding about the impact of the electronic correlations on the DSF of ions, we analyze the ion-acoustic mode dispersion, the wave-number dependence of the height and width of the ion-acoustic peak, and the ratio of  the ion-acoustic peak height to the Rayleigh peak height, i.e., $S(k,0)$.

  \begin{figure}[h!]
\includegraphics[width=0.49\textwidth]{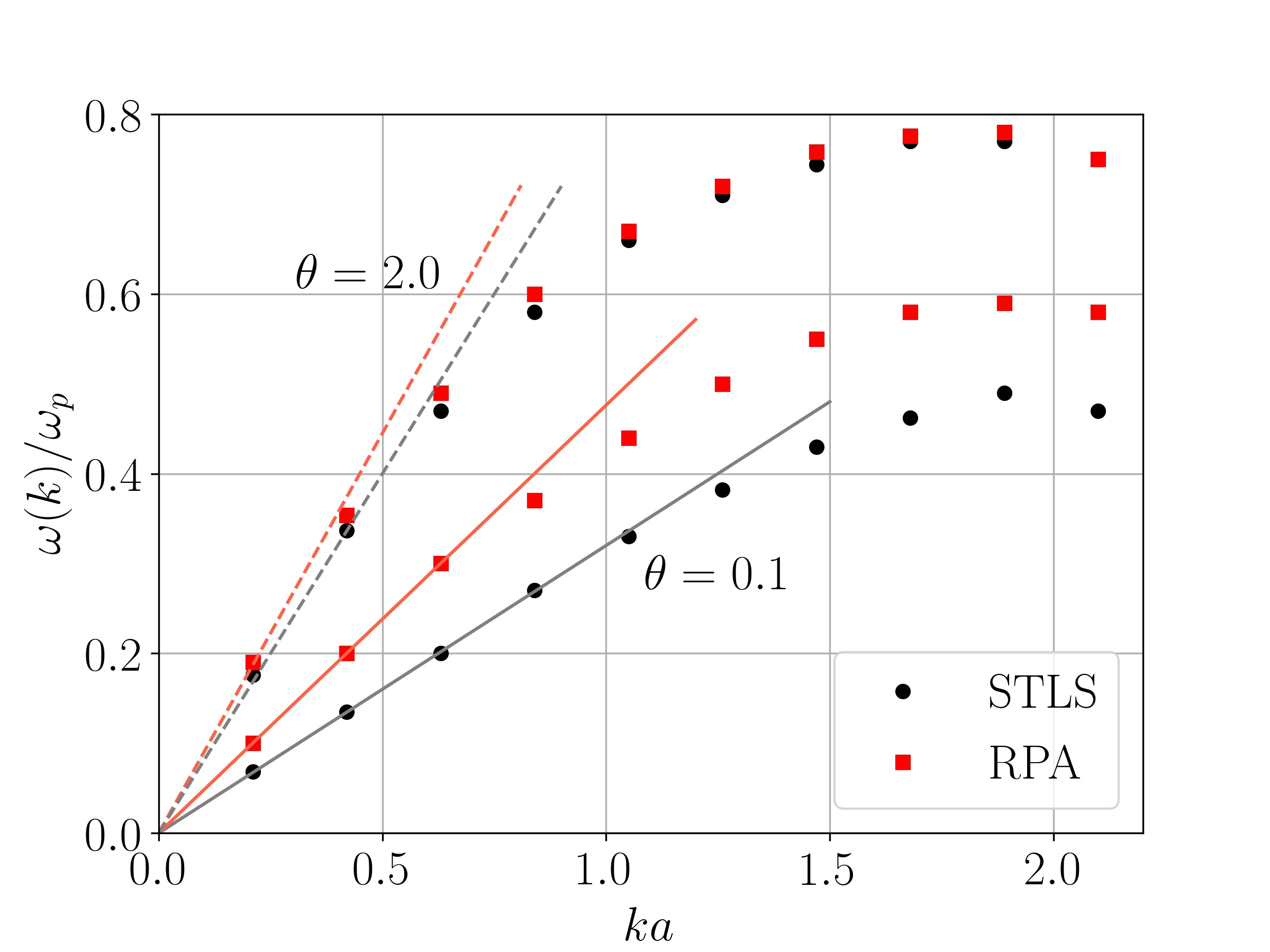}
\caption{Ion-acoustic peak positions at $r_s=1.5$ and different values of degeneracy parameter. Solid and dashed lines correspond to the linear dispersion at small wave-numbers.}
\label{fig:3}
\end{figure}

  \begin{figure}[h!]
\includegraphics[width=0.49\textwidth]{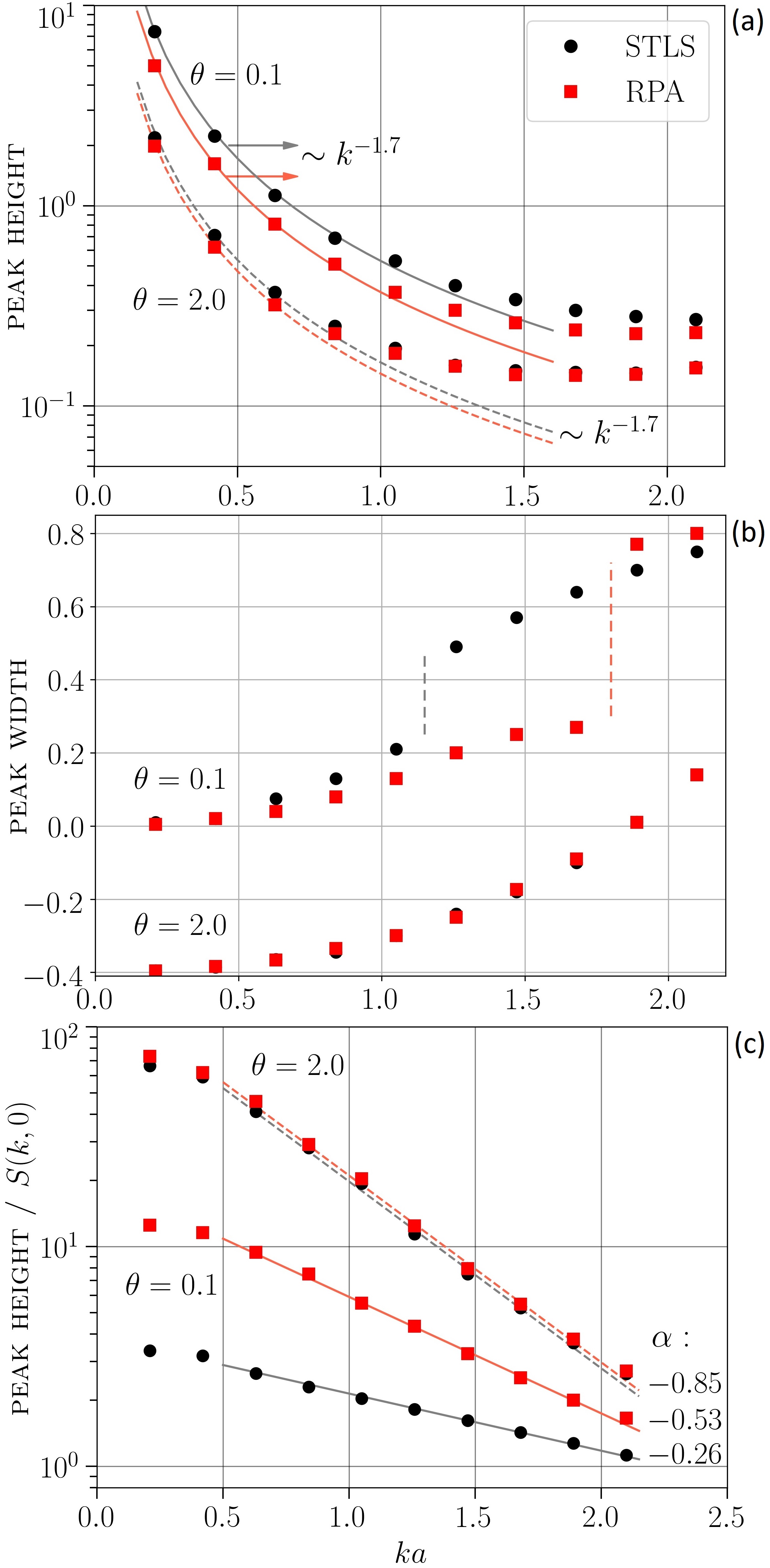}
\caption{ Ion-acoustic peak height (a) and width (b)  at $r_s=1.5$ and two values of the degeneracy parameter. The width is measured at the half height of the ion-acoustic peak. The width  values at $\theta=2.0$ are shifted by $-0.4$ for a better illustration.  The sharp increase of the ion-acoustic peak width (indicated by  a dashed vertical line) takes place when the half of ion-acoustic peak height becomes lower than $S(k,0)$. In panel (c), the ratio of  the ion-acoustic peak height to the Rayleigh peak height is shown, and the ratio of heights is described by a dependence of the form $10^{\alpha (ka)}$  at $0.5<ka<2.2$ [$\alpha$ is indicated in the legend of panel (c)].}
\label{fig:4}
\end{figure}

  \begin{figure}[h!]
\includegraphics[width=0.49\textwidth]{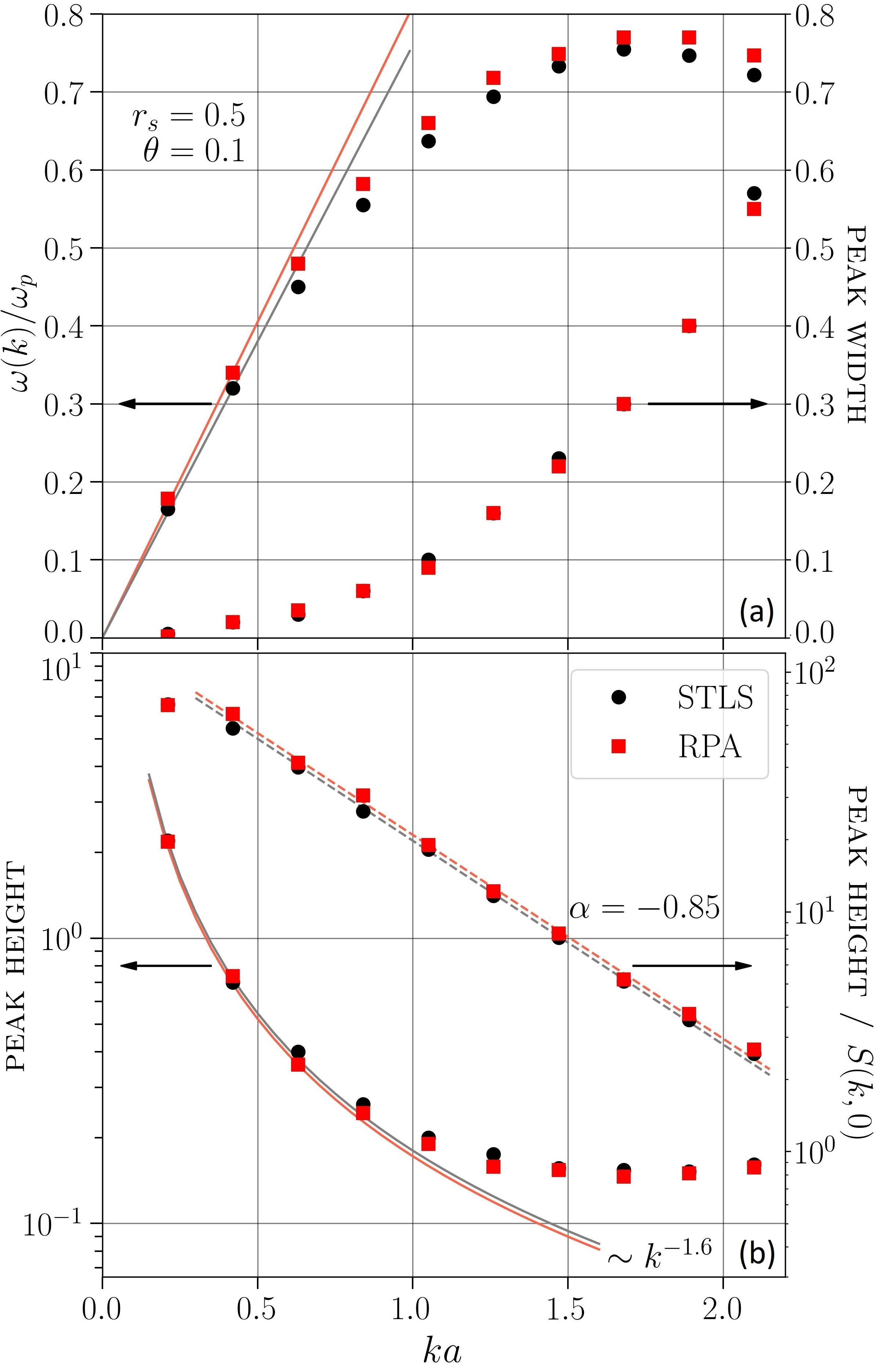}
\caption{(a) Ion-acoustic peak positions and width at $r_s=0.5$ and $\theta=0.1$, with solid lines corresponding to the linear dispersion at small wave-numbers. In panel (b), the ion-acoustic peak height and the ratio of  the ion-acoustic peak height to the Rayleigh peak height are presented; the ratio of heights is fitted by a $10^{\alpha (ka)}$ law at $0.5<ka<2.2$.}
\label{fig:5}
\end{figure}

In Fig.~\ref{fig:3}, the ion-acoustic mode dispersion is shown for $\theta=0.1$ and $\theta=2.0$ at $r_s=1.5$. From this figure, it is seen that, at $\theta=0.1$ and $r_s=1.5$, the difference between the ion-acoustic peak position taking into account electronic correlations (denoted as STLS) and that of neglecting electronic correlations  (denoted as RPA) remains significant over the entire considered range of wave-numbers, $0.21\leq ka\leq 2.21$. In contrast, at $\theta=2.0$ the difference between STLS and RPA data rapidly vanishes with increase in the wave-number. At $\theta=2.0$, the STLS and RPA data are almost indistinguishable at $ka>1.0$.  In Fig.~\ref{fig:3},  the ion-acoustic mode frequency is interpolated by a linear dependence at small wave-numbers, i.e., $\omega(k)=C_s k$. This allows us to extract the value of the ion-sound speed $C_s$ (in units of $a\,\omega_p$).
At $\theta=0.1$, for STLS and RPA data we have $C_s^{STLS}=0.32$ and 
$C_s^{RPA}=0.48$, 
respectively. This corresponds to a $33~\%$ difference due to electronic correlations. At $\theta=2.0$, we find $C_s^{STLS}=0.80$ and $C_s^{RPA}=0.89$, representing a $10 ~\%$ difference in the sound speed.  An additional interesting feature is that the approximation $\omega(k)=C_s k$ provides a remarkably accurate description of the STLS data up to $k_{\rm max}\simeq 1.2$, whereas for the RPA data, this only holds for up to $k_{\rm max}\simeq 0.6$. 
Considering Yukawa system, in Ref.~\cite{Mithen2}  it was shown that the applicability range of the hydrodynamic description is given by $k<0.43 \kappa$, where $\kappa=a/\lambda$ is the screening parameter (screening length $\lambda$).
As the electronic correlations lead to a more strongly screened potential compared to the RPA, they result in an extension of the range of applicability of the hydrodynamic description of the ionic dynamics.


The dependence of the ion-acoustic peak height on the wave number for $\theta=0.1$ and $\theta=2.0$ (at $r_s=1.5$) is shown in Fig.~\ref{fig:4} (a). As mentioned before, the peak height decreases with the  wave-number. Additionally, from  Fig.~\ref{fig:4} (a) we see that in the hydrodynamic regime (at $k<k_{\rm max}$), the peak height decay can be approximated by a power-law dependence, $k^{-\alpha}$, with $\alpha \approx 1.7$ for both STLS and RPA data at $ka>0.2$. Such a  power-law dependence was also found for Yukawa systems considering the ratio $S(k,\omega)/S(k)$  for analysis of the applicability of the generalized hydrodynamics \cite{Mithen2}. The width of the ion-acoustic peak at its half height is presented in Fig.~\ref{fig:4} (b). The width of the ion-acoustic peak increases with increase in wave-number and has a sharp rise at the wave-number corresponding to the case when the half of the ion-acoustic peak height becomes lower than $S(k, 0)$. Note that at $ka<1$, the width of the ion-acoustic peak of the STLS data is approximately the same as that of the RPA data. 

From  Fig.~\ref{fig:4} (c), we see that the ratio of the ion-acoustic peak height to the Rayleigh peak height decreases with increasing wave-number. More precisely, this ratio is characterized by an exponential decay of the form $10^{\alpha (ka)}$ in the range $0.5<ka<2.2$, and we find $-1<\alpha<0$ as indicated in Fig.~\ref{fig:4} (c).

For $r_s=0.5$ and $\theta=0.1$,
the ion-acoustic peak positions and widths are shown in Fig.~\ref{fig:5}~(a). Here, we see that the difference in the ion-acoustic peak positions of the STLS and RPA data has a small value for all wavenumbers.
The ion-acoustic peak widths, computed at the half of peak height, have almost the same value for both the STLS and RPA data. 
For the sound speed, we find $C_s^{STLS}=0.76$ and $C_s^{RPA}=0.81$, with a corresponding difference of $6~\%$. 
In Fig.~\ref{fig:5}~(b), the ion-acoustic peak height and the ratio of the ion-acoustic peak height to the Rayleigh peak height are presented. The former is described by a $k^{-1.6}$ decay at $ka<1.0$ and the latter by a $10^{-0.85 ka}$ decay at $0.5<ka<2.2$. 

\begin{figure}[h]
\includegraphics[width=0.49\textwidth]{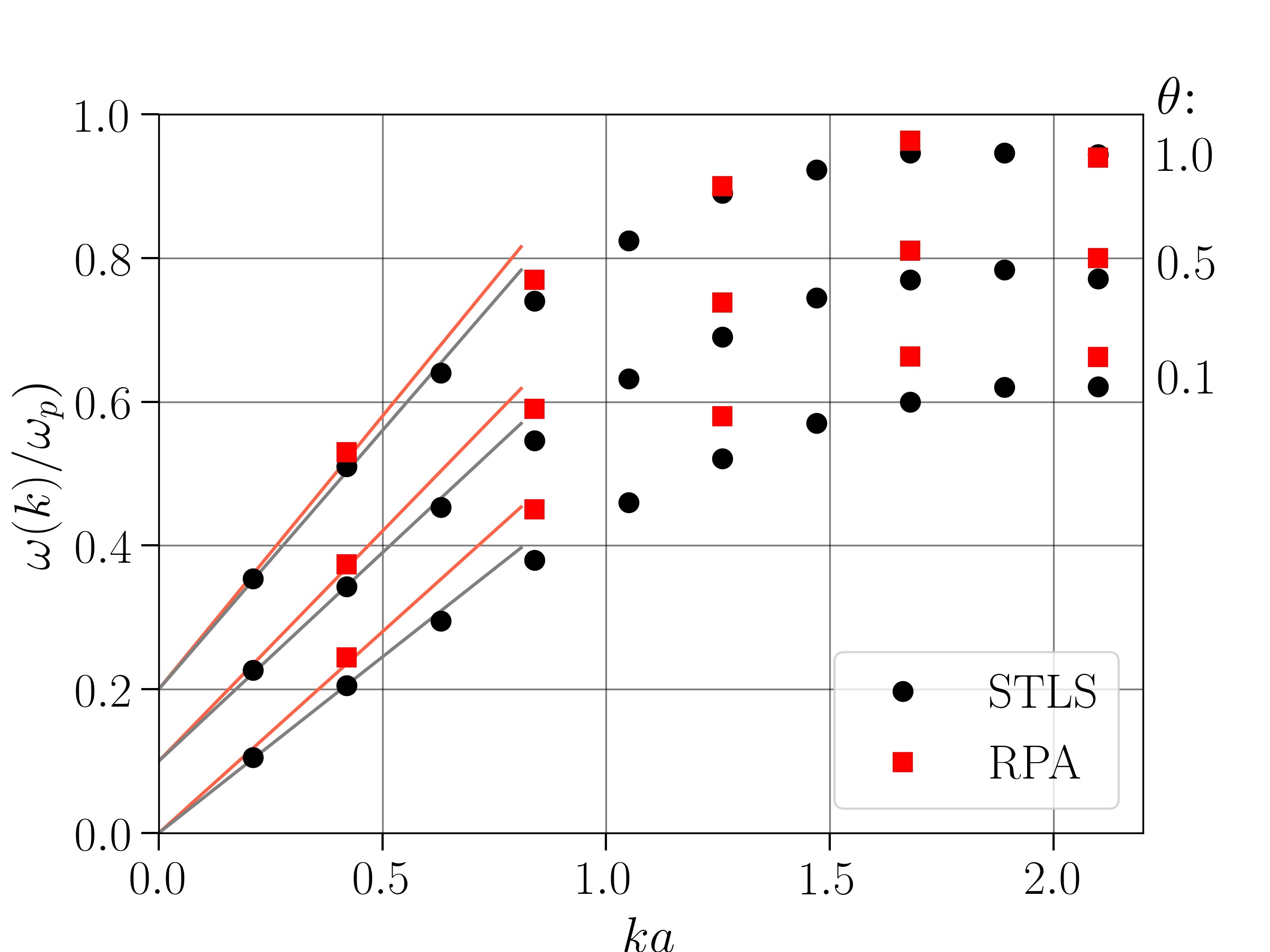}
\caption{Ion-acoustic peak positions at $r_s=1.0$ and different values of the degeneracy parameter.  The solid line corresponds to the linear dispersion at small wave-numbers. Data for $\theta=0.5$ and $\theta=1.0$ are shifted by $0.1$ and $0.2$, respectively.}
\label{fig:6}
\end{figure}

\begin{figure}[h]
\includegraphics[width=0.49\textwidth]{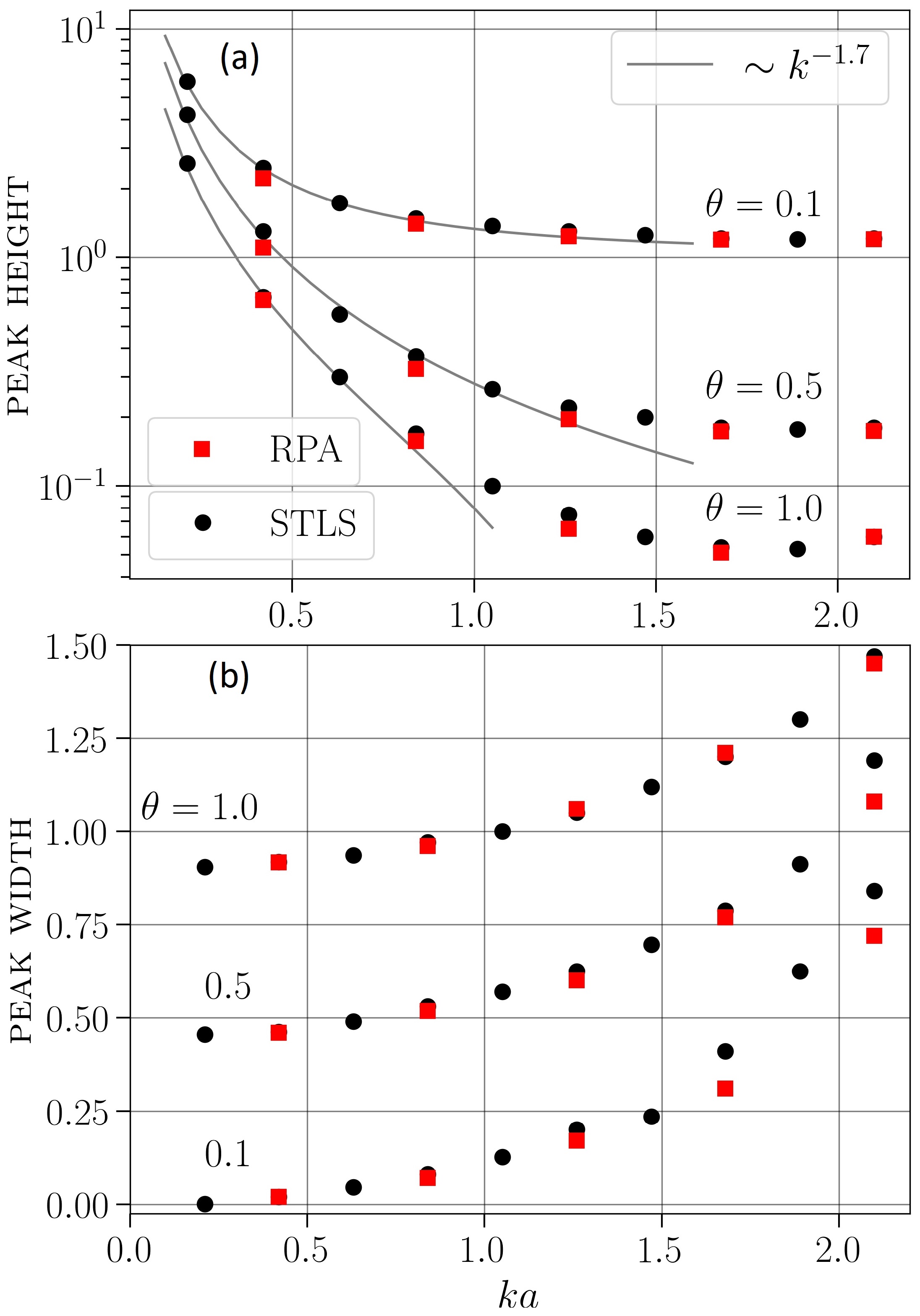}
\caption{Ion-acoustic peak height (a) and width (b) at $r_s=1.0$ and three  values of the degeneracy parameter. For better illustration, the values of the ion-acoustic peak height for $\theta=0.1$ ($\theta=1)$ are shifted by $+1$ ($-0.1$). Further, the ion-acoustic peak width  values at $\theta=0.5$ and $\theta=1.0$  are shifted by $+0.45$ and $+0.9$, respectively. }
\label{fig:7}
\end{figure}

For $r_s=1.0$ and three values of the degeneracy parameter ($\theta=0.1$, $0.5$, and $1.0$),  the ion-acoustic peak positions are presented in Fig.~\ref{fig:6}. The corresponding peak height and width dependencies on the wave-number are shown in Fig.~\ref{fig:7}.  Again, by comparing STLS and RPA data, one can see that the electronic correlations lead to lower values of the ion-acoustic mode frequencies. Upon increasing the degeneracy parameter, the difference between STLS and RPA data vanishes. In stark contrast, at $\theta=0.1$ the difference in sound speed,  which was obtained by the linear interpolation of  STLS and RPA data at small wave-numbers, is $14.3~\%$. This difference reduces to $10.3~\%$ and $5.5~\%$ at $\theta=0.5$ and $\theta=1.0$, respectively. In addition, in Fig.~\ref{fig:7} (a) we see that, at small wave numbers ($ka\leq 1$), the ion-acoustic peak height decay is described by a $k^{-1.7}$ law.  Note that at $r_s=1.0$ and the considered values of $\theta$, the ion-acoustic peak
width has approximately the same value for STLS and RPA. The ratio of  the ion-acoustic peak height to the Rayleigh peak height decays as $10^{\alpha~(ka)}$  at $0.5<ka<2.2$ (not shown in this figure), with $\alpha=-0.78$ at $\theta=1.0$, $\alpha=-0.65$ at $\theta=0.5$, and $\alpha=-0.55$ at $\theta=0.1$.

\subsection{Yukawa model with a fitting parameter}\label{s:4B}

The Yukawa model was often used for the investigation of the properties of strongly coupled ions \cite{Hansen, Donko, Mithen, Mithen2}. The corresponding dimensionless interaction potential between particles is given by
\begin{equation}\label{eq:yuk}
   \Phi(R)=\frac{\Gamma}{R}\exp(-\kappa R), 
\end{equation}
where $R=r/a$ is the distance between two ions in units of the mean inter-ionic distance, and $\kappa$ denotes the screening parameter. 

Accurate analytic results for the  dynamical structure factor of a Yukawa system were derived  on the basis of the hydrodynamic description  and the memory function formalism (see, e.g., \cite{Hansen_book, Mithen, Mithen2}). In particular, these results are handy for the analysis and interpretation of x-ray scattering data from experiments or, alternatively, from first principle simulations such as DFT. 
Therefore, achieving a better understanding of Yukawa systems is highly desirable, and the natural question arises whether STLS and RPA data can be reproduced on the basis of the ansatz~(\ref{eq:yuk}).

\begin{figure}[h!]
\includegraphics[width=0.49\textwidth]{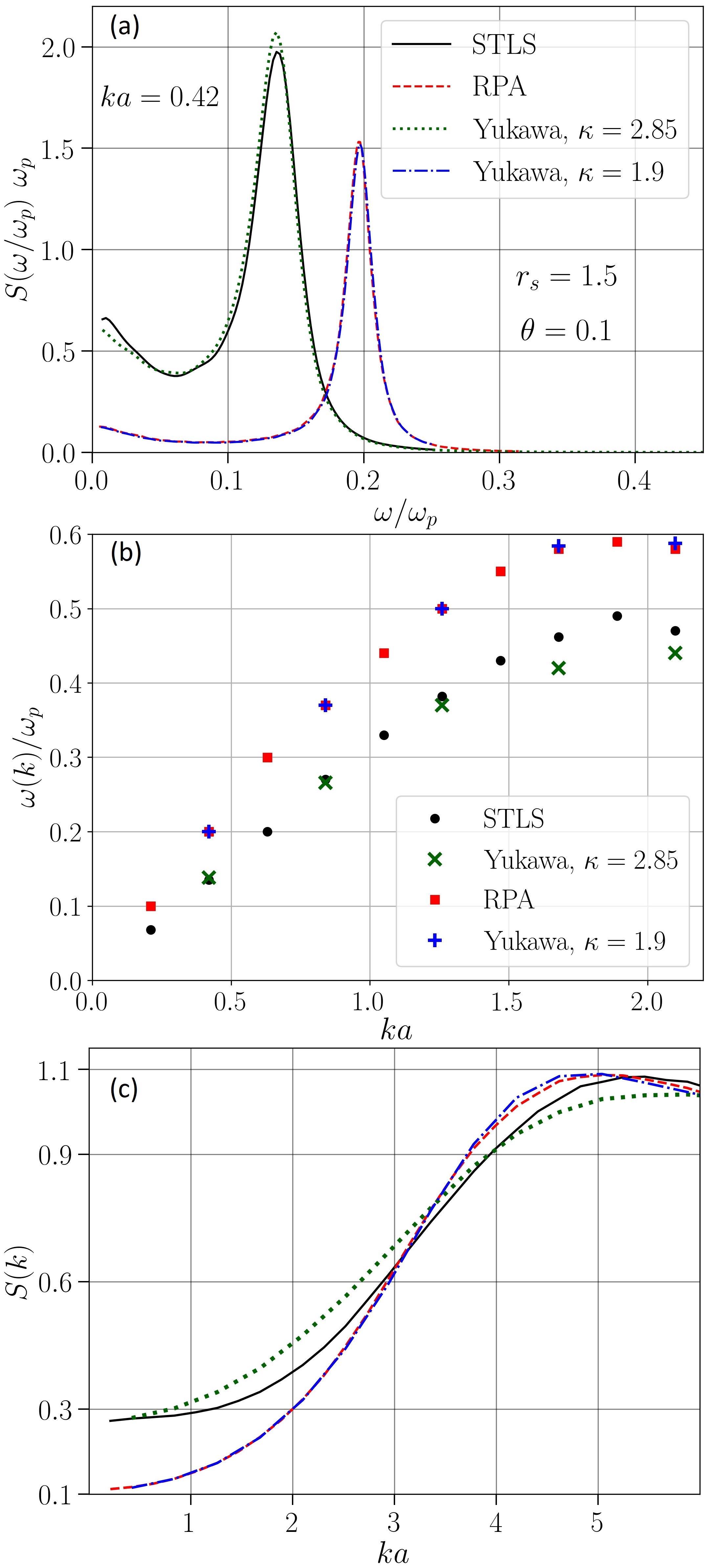}
\caption{(a) Dynamical structure factor of the ions at $r_s=1.5$ and $\theta=0.1$. (b) Ion-acoustic peak positions. (c) Static structure factor of the ions. The STLS- and RPA-potential based data are compared to the results of the Yukawa model. }
\label{fig:8}
\end{figure}

For $r_s=1.5$ and $\theta=0.1$, in Fig.~\ref{fig:8} we show (a) the ionic DSF at $ka=0.42$, (b) the ion-acoustic peak positions, and (c) the static structure factor. From Figs.~\ref{fig:8} (a) and (b), we observe that the STLS data for the DSF at $ka\leq 1.26$ and $\omega<\omega_p$ are well reproduced by a Yukawa potential with $\kappa=2.85$, where $\kappa$ should be understood as a fitting parameter.  Note that the used value of $\kappa$ provides agreement between the static structure factors of STLS and Yukawa models only for much smaller wave numbers $ka<0.5$ (see Fig.~\ref{fig:8}) (c). Therefore, the Yukawa model (with $\kappa$ used as a fitting parameter) is able to describe the DSF at $\omega<\omega_p$ in wider range of wave-numbers than the corresponding static structure factor. Essentially, the normalization of $S(k,\omega)$ (i.e., $S(k)$) can be inaccurate, while the peak position of $S(k,\omega)$ (i.e., $\omega(k)$) is still remarkably precise. 
Notably, it follows that the Yukawa model provides an accurate description of the STLS data in the hydrodynamic regime, which for $r_s=1.5$ and $\theta=0.1$ is defined by $k_{max}\simeq a^{-1}$  (see Fig.~\ref{fig:3}). 

Additionally, from Fig.~\ref{fig:8} we conclude that the RPA data is  reproduced using the Yukawa potential (with $\kappa=1.9$) at all considered values of the wave-number, where $\kappa$ is computed from the long wavelength limit of the RPA polarization function as $\kappa_Y=a\left[\frac{1}{2} k_{TF}^2 \theta ^{1/2} I_{-1/2}( \mu/k_BT_e)\right]^{1/2}$ \cite{POP15}, with $I_{-1/2}$ being the Fermi integral of order $-1/2$, and $\mu$ being determined by the normalization, $n= \sqrt{2} I_{1/2}(\mu/k_BT_e) / \pi^2(k_BT_e)^{-3/2}$.

\section{Summary} \label{s:dis}
By analyzing the DSF of strongly coupled ions, we found that electronic correlations lead to a shift of the collective ion-acoustic oscillation frequency to lower values and to higher values of both the ion-acoustic and Rayleigh peak heights. Interestingly,  the width of the  ion-acoustic peak at $ka\leq 1$ (computed at half of its height) remains the same as for the ideal case.

Remarkably, the analysis of the ion-acoustic dispersion has revealed that electronic correlations result in the extension of the applicability limit of the hydrodynamic description of the ionic dynamics with respect to larger wave-numbers and can significantly decrease the value of the sound speed (up to $30~\%$ ($10~\%$) at $r_s=1.5$ and $\theta=0.1$ ($\theta=2.0$)). We found that in plasmas with strongly degenerate electrons ($\theta=0.1$), the impact of the electronic non-ideality on the sound speed remains appreciable even in the weak electronic coupling case ($r_s=0.5$), decreasing  the sound speed value by $6~\%$ in comparison with the case neglecting electronic correlations. 

By investigating the ion-sound peak height decay with  wave-number, we showed that, at $ka\leq 1$ the ion-sound peak height decreases approximately as $k^{-n}$ with $n$ being in the range between 1.7 and 1.6 at the considered plasma parameters. Additionally, we found that the ratio of  the ion-acoustic peak height to the Rayleigh peak height (i.e. $S(k,0)$) decreases at $0.5\leq ka \leq 2.2$ as $10^{\alpha (ka)}$ with $-1<\alpha<0$. These findings are valid whether electronic correlations are included or neglected.    

Further, we demonstrated that the Yukawa model with a fitted screening parameter is able to provide an accurate description of the DSF (at $\omega<\omega_p$) and ion-acoustic mode (computed using the STLS potential) in the hydrodynamic regime ($ka<1$). This is a very important finding as previously derived analytic models for Yukawa systems in the hydrodynamic regime \cite{Mithen, Mithen2} can be used for the interpretation and analysis of experimental x-ray scattering data  \cite{McBride}.  However, the Yukawa model provides an unsatisfactory description of the static structure factor obtained using STLS potential, where the two models are in agreement  only at very small wave-numbers ($ka<0.5$).

The behavior of the DSF is most interesting at intermediate and small wave-numbers, since at large wave-numbers the ion-acoustic mode is strongly damped. From our new results, it follows that at intermediate and small wave-numbers, the reproduction of a DSF obtained experimentally (or using ab initio simulation) by using a model potential (e.g. a Yukawa potential) does not require full agreement with respect to the static structure factor.
Moreover, model potentials that are designed to accurately reproduce the static structure factor can lead to incorrect and misleading results for dynamical properties \cite{Harbour}. 

On a final note, we mention that Cl\'erouin et al \cite{Clerouin2} have shown that the Yukawa model can reproduce OFMD results for the ionic DSF at $ka<0.5$. To have such an agreement they fitted the value of the effective ionic charge and used the screening parameter of ideal electrons. However, from our new results presented in Sec.~\ref{s:4B}, we conclude that a much better strategy is given by finding a suitable screening parameter instead of using the ideal value.

 \section*{Acknowledgments}
Zh.A. Moldabekov gratefully acknowledges funding from
the German Academic Exchange Service (DAAD) under program number 57214224.   This work has been supported by 
   the Ministry of Education and Science of Kazakhstan under Grant No. BR05236730, “Investigation of fundamental problems of
physics of plasmas and plasma-like media” (2019) and by the DFG via grant BO1366/13.

%

\end{document}